\documentstyle[twocolumn,aps,psfig,epsf]{revtex}

\begin{document}
\twocolumn[\hsize\textwidth\columnwidth\hsize\csname 
 @twocolumnfalse\endcsname
 \title{ Tangled Nature: a model of evolutionary ecology.} 
\author{ Kim Christensen$^\dag$, Simone A. di Collobiano$^\dag$,
Matt Hall$^\ddag$, and Henrik J. Jensen${^\ddag *}$}
\pagestyle{myheadings}
\address{ $^\dag$Blackett
Laboratory, Imperial College, Prince Consort Road, London SW7 2BW,
U.K.\\ $^\ddag$Department of Mathematics, Imperial College, 180
Queen's Gate, London SW7 2BZ, U.K.}

\maketitle
\date{\today}
\begin{abstract}
We discuss a simple model of co-evolution. In order to emphasise
the effect of interaction between individuals the entire population is
subjected to the same physical environment. Species are emergent
structures and extinction, origination and diversity are entirely a
consequence of co-evolutionary interaction between individuals. For
comparison we consider both asexual and sexually reproducing populations.
We also study competition between asexual and sexual reproduction in a mixed population. In either case the system evolves through periods of hectic 
reorganisation separated by periods of coherent stable coexistence.
\\
$^*$Author 
to whom correspondence should be addressed. E-mail: h.jensen@ic.ac.uk

 \end{abstract}
\vskip1pc
]

\section{Introduction}
It is difficult in experiments and observations to bridge the gap between
ecological time and evolutionary  time (Pimm, 1991). 
Nevertheless, since Darwin's
publication of {\it The Origin of Species} (Darwin, 1859)
it has been generally agreed that
the intricate and complex ecologies surrounding us are the product of 
Natural Selection operating on vast numbers of successive generations.
We know that the slow gradual effect of mutations and Natural Selection
is the long term mechanism
 underlying evolution in  ecological systems, but we are often unable
to answer questions concerning stability and the nature of the dynamical
evolution (intermittent versus gradual). It is also difficult to measure
the degree of interrelatedness of an ecology (e.g. Bj\o rnstad {\it et al.}
, 2001): who is 
interacting with whom and how strongly, and it is difficult to determine the species  abundance in detail. What is especially difficult is to monitor the 
temporal  variation  in the aforementioned quantities over evolutionary time. 

Can general principles be identified for the overall dynamical
behaviour of evolution? Even if the characteristics  of
each  individual species have to be considered in their proper
specific contexts, perhaps general laws do operate at an overall
 level. Obviously the answers to these questions are 
 empirical, but indicators may be obtainable from
deliberately simplified theoretical models. It is obviously important
to consider carefully the type of simplification assumed. Simplified
models often operate directly at species level and typically consider only a few
coupled species (see e.g. Doncaster, 2000) but in order to
capture the consequences of the  complexity characteristic of ecology
we believe it is important to treat species as emergent structures and
to allow for the multitude of interactions each individual
(and therefore species) is subject to. We find that the very nature of the
dynamics of the ecology is strongly influenced by the complexity
of the system.

Temporal as well as spatial variations in the physical environment are
known to play an important role in evolution. It is also often assumed
that co-evolution with interaction between co-existing individuals, or
species, may influence the evolutionary dynamics in a significant way
(Kauffman, 1995; Bak \& Sneppen, 1993). The relative importance of
selective pressure of purely physical origins and co-evolutionary 
effects is not clear, and it seems difficult to resolve the issue
solely by use of selected specific case studies. Moreover, seen from
an ecological point of view, the biotic and the physical environment
are coupled.

In the present paper we present an individual based mathematical model of 
an evolving ecology. The model is kept sufficiently simple to allow us to simulate 
evolutionary time scales. 

We attempt here to gain some
insight into the possible effects of co-evolution through the study of a
model in which variations in the physical environment are altogether
neglected. Our model is not meant to be a realistic representation of
biological evolution, but rather a theoretical approximation in which
co-evolution is made to be the prominent driving force. We then
demonstrate within this model that speciation {\it does} occur and we
study in some detail the dynamical features of the evolution of the
model as well as the nature of the ecology created by the
co-evolutionary dynamics. 

We are interested in the qualitative behaviour of a system in which
the mutual interaction between co-existing individuals of different 
genetic composition determines the possibility of the individual to
thrive. The model emphasises the web of interactions between individuals
of different genomic composition, to stress this aspect we will
talk about the Tangled Nature model - or the TaNa model for short.
We represent biotic factors in terms of
the co-evolutionary effects on the fitness of individuals. 
The model is a simplification. No distinction is made between
genotype and  phenotype and  the details of the  reproductive
mechanism are kept to a minimum. This simplification allows us to
represent evolution in terms of the dynamics of the distribution of
the population in genome space. We demonstrate that at a qualitative
level the complex dynamics of the model resembles known aspects
of long term biological evolution such as speciation and
intermittent behaviour. We are also able to study the competition between
asexual and sexual reproduction. We find that asexual reproduction is
superior during periods of rapid relocation of the configuration in genome
space, whereas sexual reproduction is most advantageous during the 
coherent more stable epochs. It is natural to relate these stable
epochs to periods of Evolutionary Stable Strategies (ESS), as introduced
by Maynard Smith (1982). The stable periods of our model are,
however, not perfectly stable as fluctuations caused by mutations
can trigger a switch from one stable period to another. We therefore
suggest calling these periods ``quasi-Evolutionary Stable Strategies''  
or q-ESS. The overall effect of the evolutionary dynamics of 
the present model is {\em to increase the average duration of the q-ESS}.

\section{Definition of the Model} 

\noindent {\bf Interaction}\\
We now define the  TaNa model in detail.
We represent individuals in the same way as in
models considered by, e.g., Kauffman (1995), Higgs and Derrida (1992), 
Gavrilets (1999), Eigen {\it et al.} (1988), and Wagner {\it et al.}
 (1998).
 An individual is
represented by a vector ${\bf S}^\alpha=
(S_1^\alpha,S_2^\alpha,...,S_L^\alpha)$ in genotype space.
Here $S^\alpha_i$ may take the values $\pm 1$. These may be
interpreted as genes with two alleles, or a string of either
pyramidines or purines. Individuals are labelled by Greek letters 
$\alpha,\beta, ... = 1,2,..., N(t)$. 
When we refer, without reference to a specific individual,
to one of the $2^L$ positions in genome space,
we use roman superscripts ${\bf S}^a$, ${\bf S}^b$, ... with 
$a,b,... =1,2,...,2^L$. Note, many different individuals 
${\bf S}^\alpha, {\bf S}^\beta$,. . ., may reside on the same
position, say ${\bf S}^a$ in genome space.Geometrically the vector  ${\bf S}^a$
represents one of the corners of the $L$ dimensional hyper-cube ${\cal
S}=\{-1,1\}^L$ (see Fig. 1). 

\begin{figure}[ht]
\vspace{0cm}
\centerline{\hspace{0cm}
\psfig{figure=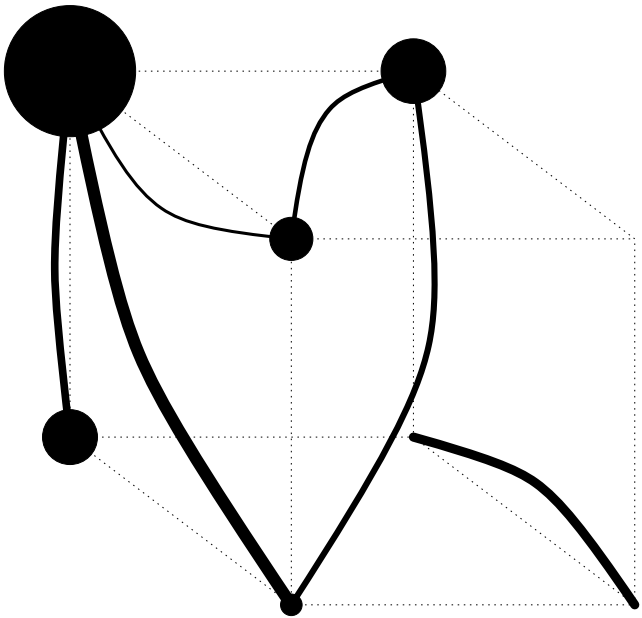,width=8.5cm,angle=-90}}
\vspace{0cm}
\caption{ 
A three dimensional genome space. For $L=3$ the sequence of genes uniquely
defines a vertex of a cube. The number of edges (dotted lines) that must
be traversed between 2 vertices defines their Hamming distance.
Interactions between vertexes are shown as solid curves with
thickness indicating the strength and circles placed at the vertexes
have radii proportional to the occupation (number of individuals
present) with the genome in question. Note that interactions are defined
even for unoccupied vertices. } 
\label{1}
\end{figure}

The ability of an individual $\alpha$ to reproduce is
controlled by $H({\bf S}^\alpha,t)$:
\begin{equation}
H({\bf S}^\alpha,t)={1\over N(t)}\sum_{\beta=1}^{N(t)} \sum_{i=1}^L
J_i({\bf S}^\alpha,{\bf S}^\beta)S_i^\alpha S_i^\beta-\mu N(t), 
\label{Hamilton1}
\end{equation}
where $N(t)$ is the total number of individuals at time $t$. The sum
over individuals $\beta$ in Eq. (\ref{Hamilton1}) is more conveniently
expressed as a sum over the locations ${\bf S}$ in the genome space
${\cal S}$, using the occupancy $n({\bf S},t)$ of the locations we obtain:
\begin{equation}
H({\bf S}^\alpha,t)={1\over N(t)}\sum_{{\bf S}\in{\cal S}}n({\bf S},t)
\sum_{i=1}^L J_i({\bf S}^\alpha,{\bf S})S_i^\alpha S_i
-\mu N(t).
\label{Hamilton2} 
\end{equation}
Two positions ${\bf S}^a$ and ${\bf S}^b$ in genome space are coupled
 with the fixed strength ${\bf J}^{ab}={\bf J}({\bf S}^a,{\bf S}^b)$. This
 coupling is non-zero with  probability $\Theta$, in which case 
 we assume ${\bf J}^{ab}\neq {\bf J}^{ba}$ to be random and uniform
on the interval $[-c,c]$, where $c$ is a constant.
The structure of the coupling in genome space is sketched in Fig. 1.

Some comments about the
interaction matrix ${\bf J}({\bf S}^a,{\bf S}^b)$ are appropriate. 
In our simplistic approach a given genome is imagined to lead uniquely
to a certain set of attributes (phenotype) of the individuals/organisms.
The locations ${\bf S}^a$ and ${\bf S}^b$ represent blueprints for
organisms that exist {\it in potentia}. The locations may
very likely be unoccupied but, if we were to construct individuals
according to the sequences   ${\bf S}^a$ and ${\bf S}^b$ the two
individuals would have some specific features. Anecdotally we can
imagine that ${\bf S}^a$ corresponds to rabbits and ${\bf S}^b$ represents
 foxes. The number ${\bf J}({\bf S}^a,{\bf S}^b)$ now represents the potential 
influence of an individual constructed according to the genome sequence 
${\bf S}^b$ on an individual constructed according to the genome sequence 
${\bf S}^a$. In our toy example ${\bf J}({\bf S}^a,{\bf S}^b)$ represents
the fact that the foxes will tend to eat the rabbits and thereby
decrease the rabbits ability to survive and ${\bf J}({\bf S}^b,{\bf S}^a)$
represents the fact that the availability of rabbits as a food source
will help to sustain the foxes. Other examples could be parasitic 
or collaborative relationships.
In order to emphasis co-evolutionary aspects we have {\em excluded}
``self-interaction'' among individuals located at the same positions
${\bf S}$ in genome space, i.e., we use ${\bf J}({\bf S},{\bf S})={\bf 0}$.

In reality the mutual influence between two individuals of a
certain genotype (phenotype) is, of course, not a random quantity.
The interaction maybe be collaborative,
competitive or neutral. It is this aspect 
we represent by ascribing a set of fixed  randomly assigned
coupling strengths between the positions in genome space.

We stress that the segregation (or speciation) to be
discussed below is an effect of different couplings
between different positions ${\bf S}^a$ and ${\bf S}^b$. When
we assume $J_i({\bf S}^a,{\bf S}^b)=J_0$ independent of
${\bf S}^a$ and ${\bf S}^b$, we find the population not to
be concentrated around a subset of all positions
in genome space, instead the population is smeared
out through the genome space in a diffuse manner. 

The conditions of the physical environment are simplistically
 described by the term $\mu N(t)$ in Eq. (1), 
where $\mu$ determines the average 
sustainable total population size. An increase in $\mu$ corresponds
 to more harsh physical conditions. This is a simplification,
though one should remember that the physical environment encountered
by an organism is to some extent produced by the presence
of other living organisms. Consider, for example, the environment experienced
by the bacterial flora in the intestines. Here one type of bacteria very much
live in an environment strongly influenced by the presence of other types
of bacteria. In this sense some fluctuations in the environment may
be thought of as included in the matrix ${\bf J}({\bf S}^a,{\bf S}^b)$.

\noindent{\bf Reproduction}\\
 Asexual reproduction consists of one individual
being replaced by two copies, this event occurs for individuals
${\bf S}^\alpha$ with 
a probability per time unit proportional to
\begin{equation}
p_{off}({\bf S}^\alpha,t)={ \exp[H({\bf S}^\alpha,t)]\over
1+\exp[H({\bf S}^\alpha,t)]}\in[0,1].
\label{p_off}
\end{equation}

In the case of sexual reproduction an individual ${\bf S}^\alpha$ is
 picked at random and paired with another randomly chosen individual
 ${\bf S}^\beta$ with Hamming distance
 $d={1\over2}\sum_{i=1}^L|S_i^\alpha-S_i^\beta|\leq d_{max}$ (allowing at
 most $d_{max}$ pairs of genes to differ). The pair produces an offspring
 $\gamma$ with a probability $\sqrt{p_{off}({\bf
 S}^\alpha,t)p_{off}({\bf S}^\beta,t)}$, with $S_i^\gamma$ chosen at
 random from one of the two parent genes, either $S_i^\alpha$ or $S_i^\beta$.
 
\noindent {\bf Mutation}\\
 Genes mutate with probability $p_{mut}$, represented
by a change of sign $S_i^\gamma \rightarrow - S_i^\gamma$, during the
reproduction process. Choosing genes at random from the parents
may be thought of as a process similar to
recombination for $d_{max}\geq2$.

\noindent{\bf Annihilation}\\
For simplicity an individual is removed from
the system with a constant probability $p_{kill}$ per time step.
This procedure is implemented both for asexual and sexually reproducing
individuals. 

\noindent{\bf Time Step}\\
A time step consist of {\bf one} annihilation attempt followed
by one reproduction attempt. One generation consists of $N(t)/p_{kill}$
time steps, the average time taken to kill all currently living individuals.

\noindent{\bf  Stability}\\
 At an average level of description, and neglecting mutations, the
above dynamics is described by the following set of equations (one
equation for each position  in the genotype space):
\begin{equation}
{\partial n({\bf S},t)\over \partial t} = [p_{off}({\bf S},
t)-p_{kill}]n({\bf S},t)
\label{mean_field}
\end{equation}                            
controlling the temporal evolution of the occupancy $n({\bf S},t)$ of
the positions ${\bf S}$ in genotype space ${\cal S}$. Stationary
solutions (i.e., those for which $\partial n/\partial t=0$)
 demand either $n({\bf S},t)=0$ or $p_{off}({\bf S},t)=
p_{kill}$.  During the q-ESS the system manage to find a
configuration in genotype space for which all occupied positions satisfy
the balance between production of offspring and decease. The fitness
$p_{off}({\bf S}^a,t)$ of individuals at a position   ${\bf S}^a$
depends on the occupancy $n({\bf S}^b,t)$  of all the sites  ${\bf
S}^b$ with which site ${\bf S}^a$ is connected through couplings
${\bf J}^{ab}$. Accordingly, a small perturbation in the occupancy at
one position is able to disturb the balance between $p_{off}({\bf
S},t)$ and $p_{kill}$  on connected sites. In this way an imbalance at
one site can spread as a chain reaction
through the system, possibly affecting a global reconfiguration of the
genotypical composition of the population.

\section{Dynamical behaviour}
We consider three different types of populations. (1)
a purely asexual population, (2)purely sexual 
population and (3) a mixed population in which 
mutations can transform an asexual individual into a sexually
reproducing individual and vice versa. 

\subsection{Asexual reproduction}
In this subsection we discuss the model when all reproduction is 
assumed to be asexual.

\noindent{\bf Initiation}\\
Let us consider the initiation of the model. First
we place the entire population $N(0)$ at a randomly chosen location 
${\bf S}^*$ in genome space. The $H$-function in Eq. (\ref{Hamilton2}) will be
given by $H({\bf S}^*,0)=-\mu N(0)$ since $n({\bf S})=0$ for
 ${\bf S}\neq {\bf S}^*$ and ${\bf J}({\bf S}^*,{\bf S}^*)=0$. 
If no mutations can occur the population will
remain confined at the location ${\bf S}^*$ and the size
of the population $n({\bf S}^*,t)$ will 
according to Eq. (\ref{mean_field}) approach the value 
\[ N^*={1\over \mu}\ln\left(1-p_{kill}\over p_{kill}\right). \]

Mutations do occur, however, and the population will migrate away from
the original location ${\bf S}^*$ into the surrounding 
region of genome space. In Fig. 2, we show a cladogram
indicating the evolution of the first 110 generations.
During this initial period the newly invaded positions are only
occupied for a few generations. After this period of
rapid changes a relatively stable configuration is achieved,
and the occupied positions to the right in Fig. 2 indicate that the
system has entered its first q-ESS.

\begin{figure}[ht]
\vspace{0cm}
\centerline{\hspace{0cm}
\psfig{figure=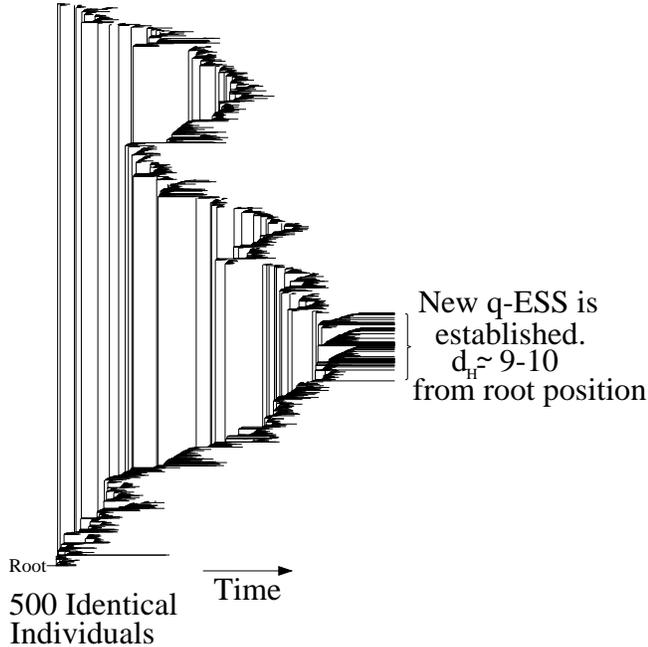,width=8.5cm,angle=0}}
\vspace{.5cm}
\caption{ 
The initial diversification from a single position in genome
space. The system is initialised at time $t=0$ with 500 identical
individuals and allowed to develop autonomously. The
system to mutates away from the initial location, which
becomes extinct relatively quickly. After 34 branchings the system
finds a stable configuration and enters the first q-ESS 
(see Figs. 3 and 4).
} 
\label{2}
\end{figure}

We have also studied simulations started out from an initial population
spread out over many randomly chosen positions in genome space.
Most of these initially occupied positions rapidly become extinct.
In this way, the diversity in genome space passes through a ``bottleneck''
before the population starts to migrate out into genome space from
one or a few positions which were able to pass through the bottleneck.
From then on, the evolution of the ecology behaves in the same way
as when started out from one single position in genome space.
 
\noindent{\bf Long time behaviour}\\
Now we turn to a discussion of the nature of the long time
dynamics of the model.
The model consists of a variable number of co-evolving individuals all
subject to the same physical environment. An individual's ability
to thrive depends on its own genetic composition as well as the
genetic composition of the other individuals present. The dynamical
evolution, driven by mutations, will have to strike a balance between
the multiplication of the individuals and the total carrying capacity
of the environment. Different types of genotypical compositions of
the population can achieve this balance. 

One possibility consists of
very numerous populations distributed on a relatively small number of
isolated regions in genotype space corresponding to a small
number of species (compare to the total number of genotypes for a given
genome length). These configurations can be stable for very many
generations and allow the species to co-exist quietly during coherent
periods of little variation in the total size or composition
 of the population, see Fig. 3.

\begin{figure}[ht]
\vspace{0cm}
\centerline{\hspace{0cm}
\psfig{figure=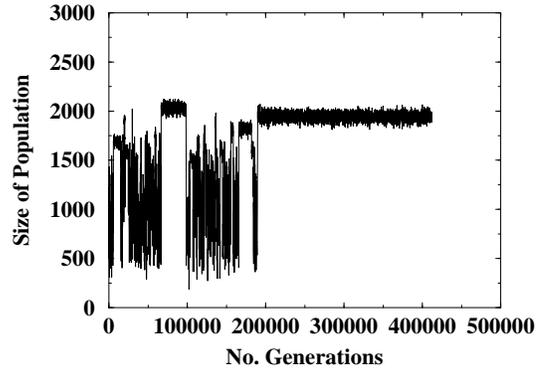,width=7.cm,angle=-90}}
\vspace{0cm}
\caption{ 
Change of total population with time for a system with
 $L=20$, $\mu=0.1375$, $p_{kill}=0.2$, 
$p_{mutate}=0.01$, $c=100$ and $\theta=0.25$. 
Regions of high population and low
 relative fluctuations (q-ESS) are
clearly distinct from regions of low population and high relative
fluctuations (transition periods). 
} 
\label{3}
\end{figure}

In Fig. 4, we demonstrate that the occupancy of the positions in genome space
fluctuates only very little during these stable periods.
We call these epochs q-ESS, or quasi Evolutionary Stable 
Strategies (Maynard Smith, 1982). The q-ESS exhibit a degree of
stability against mutation induced changes, but fluctuations in the
frequency distribution in genome space can abruptly destabilize such
a configuration.

\begin{figure}[ht]
\vspace{0cm}
\centerline{\hspace{0cm}
\psfig{figure=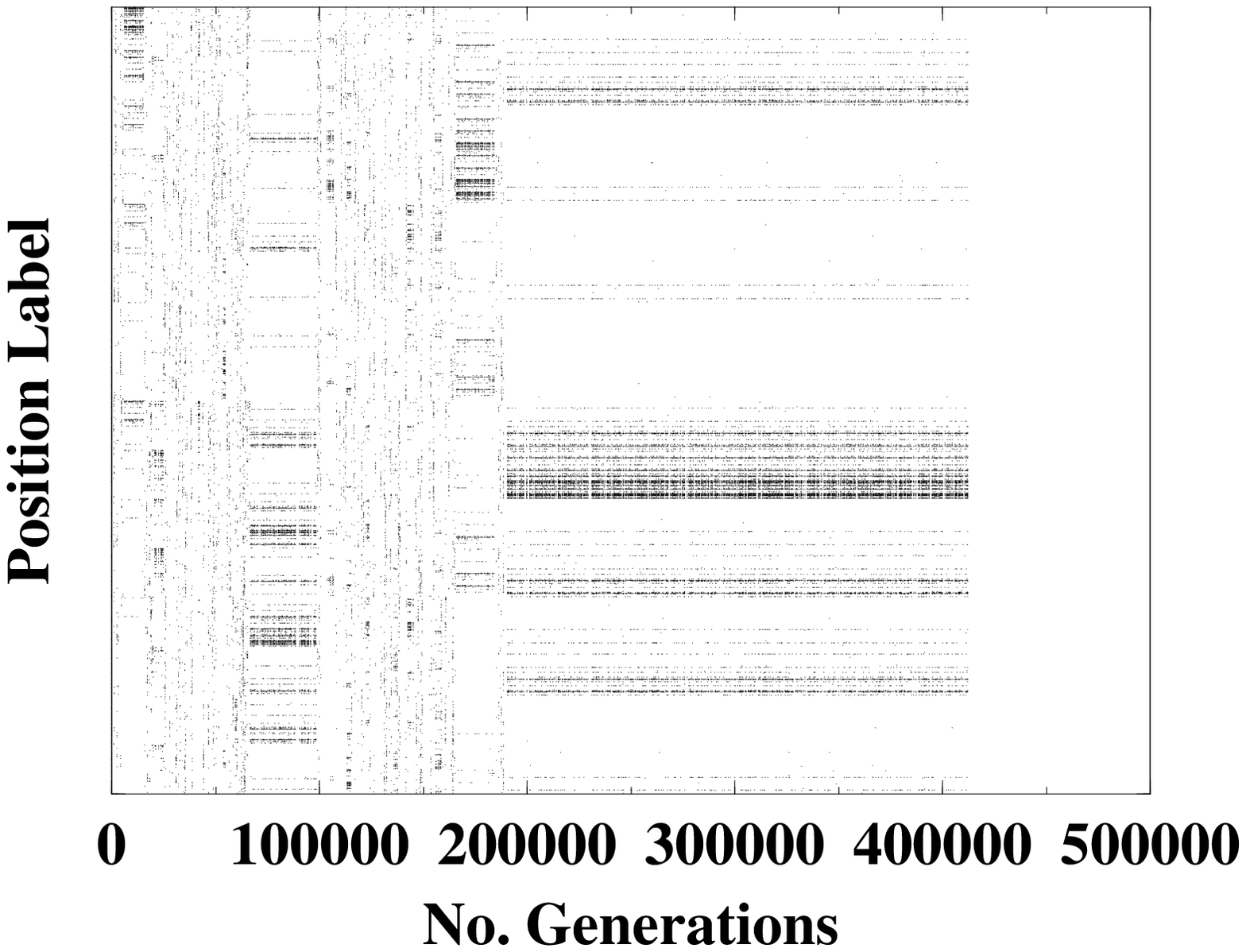,width=6cm,angle=-90}}
\vspace{2cm}
\caption{ 
Occupation of genome space versus time for the same simulation as in
Fig. 3. We arrange the positions in genome
space in a convenient arbitrary way along the y-axis and
place a dot for each occupied location at a given time.
Periods of stability (q-ESS) interrupted by periods of hectic
rearrangement are clearly visible.
} 
\label{4}
\end{figure}

We show, however,  in Fig. 5 that the distribution of
lifetimes of the q-ESS, measured in numbers
 of generations, is very broad. 

\begin{figure}[ht]
\vspace{0cm}
\centerline{\hspace{0cm}
\psfig{figure=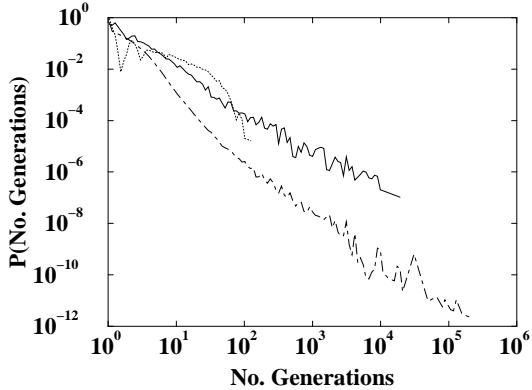,width=7cm,angle=-90}}
\vspace{0cm}
\caption{ 
Log-Log plot of the distributions of lengths of q-ESS
(solid line), transition periods (dotted line) and lifetimes of occupied
locations in genome space (dot-dashed line). We observe power-law-like
behaviour in both the q-ESS and the lifetimes of genome space
locations, but the transition periods exhibit an abrupt cut-off at much
shorter times. The lifetimes curve extends further than that of the
q-ESS, indicating that locations may remain occupied from one
q-ESS to another, surviving the transition.
} 
\label{5}
\end{figure}

\noindent{\bf Transitions}\\
The q-ESS periods are separated by periods of hectic rearrangements
of the genotypical composition of the entire population.
During these periods of rapid change, the total number of individuals
is small and populations located at specific positions in genome
space undergo sequences of bifurcations as seen in Fig. 6,
where we follow the evolution across the hectic transition period from one  q-ESS to the next.

\begin{figure}[ht]
\vspace{0cm}
\centerline{\hspace{0cm}
\psfig{figure=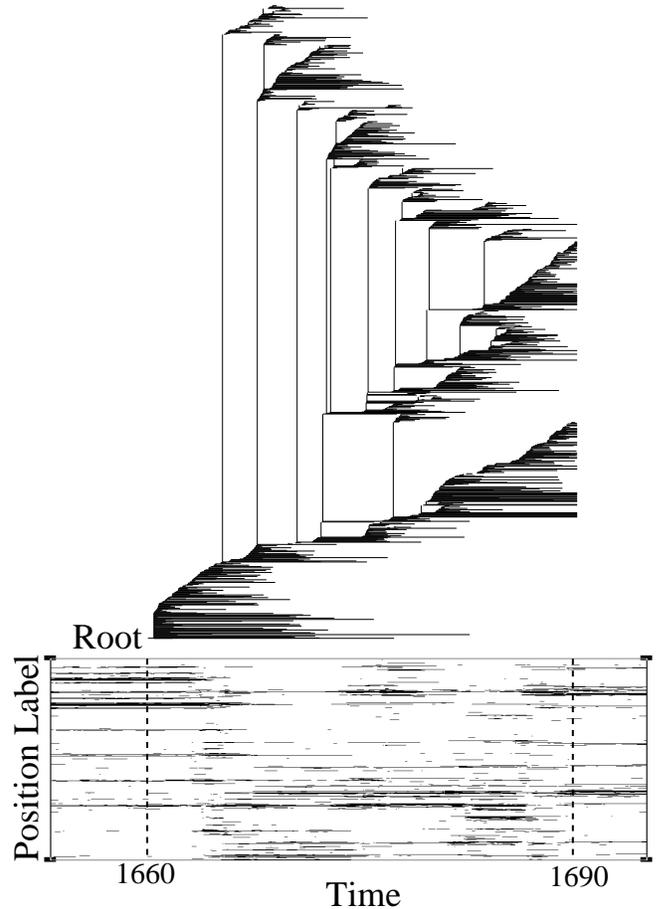,width=8.5cm,angle=0}}
\vspace{0cm}
\caption{ 
The descendants of a fit location across a transition period. The
Lower portion of this figure is a slice from an occupation plot, similar to 
Fig. 4. We track the descendants of a single, fit location across the transiton
and into the next q-ESS. After 12 Branchings the descendants have found a
new fit configuration that is stable enough to form part of the new q-ESS.
The original location does not survive the transition and lineages of the
other fit locations from the original state die out very rapidly.
} 
\label{6}
\end{figure}

The figure is a cladogram tracing out all the descendants originating 
from one root. One notices that most of the new positions spun off from 
the root die before the next q-ESS is reached.
While new branches are created old ones die. 
The periods of rapid rearrangement in genome space
are transition periods during which the 
system searches for a new stable configuration.
\newpage
.

\begin{figure}[p]
\vspace{5cm}
\centerline{\hspace{-11cm}
\psfig{figure=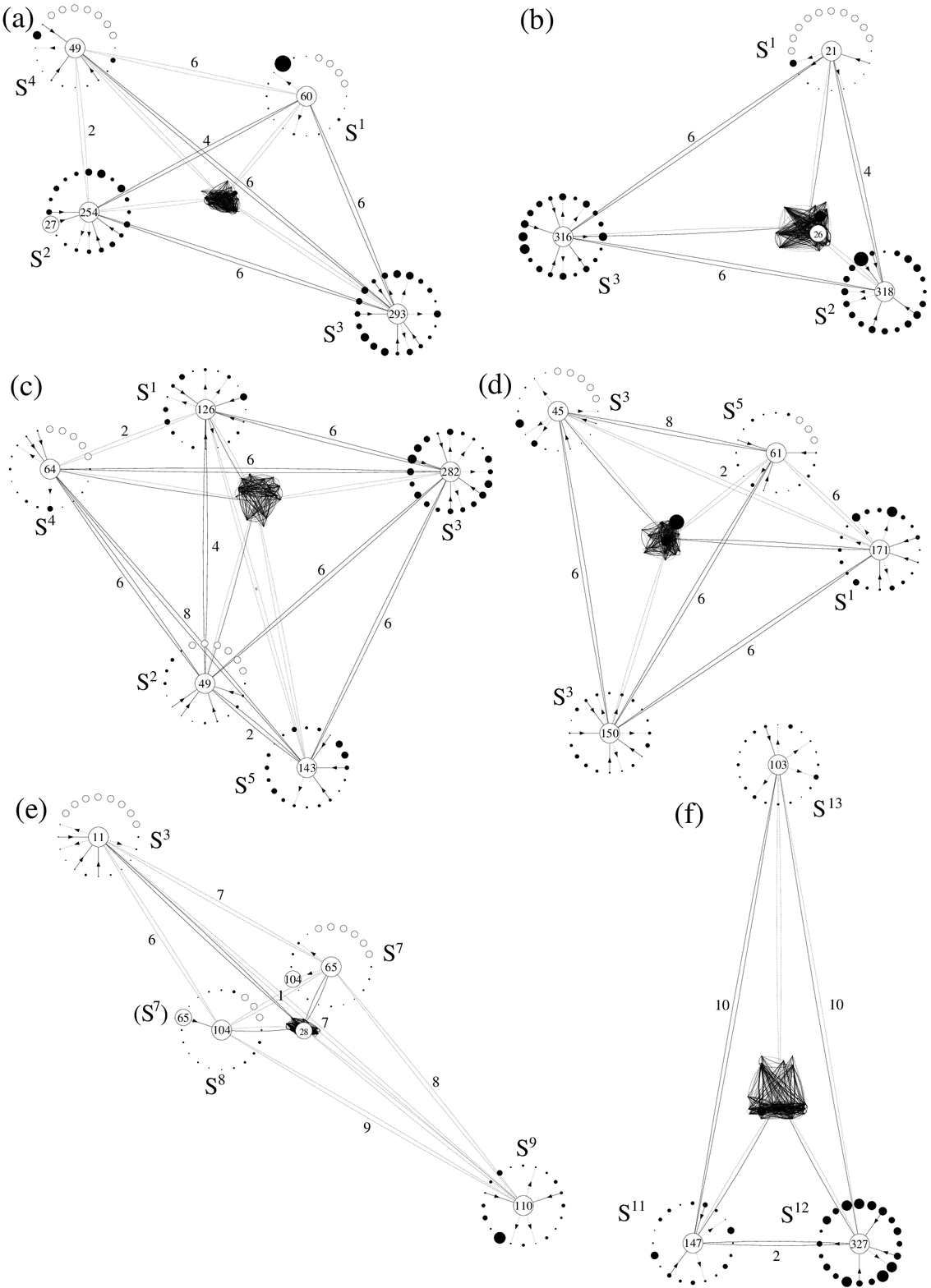,width=15cm,angle=0}}

\newpage
\caption{
Flower diagrams of the configuration of a system during a
transition from one coherent state to another. Flower diagrams visualise
the interactions and genome space proximities in a system at a given
time. Very fit locations have a complete set of nearest mutational
neighbours (these make up a "flower"). Each occupied location is
represented by a circle of radius proportional to its occupation,
(or a number, for very large locations) positive interaction 
strengths are
solid lines, and negative interaction strengths dotted lines. If a
flower is out competed by a new mutant, the q-ESS is
disturbed and the system executes an adaptive walk, searching for a new 
q-ESS. Each diagram is separated by approximately 5 to 10 generations.
See the main text for discussion. The Hamming distance between two highly
occupied positions appears midway between them.
} 
\label{7}
\end{figure}
 
The nature  of the transition from one q-ESS to the next 
is indicated in Fig. 7. This set of diagrams represent
in a quantitative way the positions with the largest occupation
together with the couplings in genome space.

Gene sequences in the diagrams are as follows:
\begin{eqnarray}
{\bf S}^1 &=&  (+1,+1,+1,-1,-1,+1,-1,+1,-1,-1,+1,-1,+1,-1,+1,-1,-1,-1,+1,-1)  \nonumber \\
{\bf S}^2 &=&  (+1,+1,-1,-1,-1,+1,-1,+1,+1,+1,-1,-1,+1,-1,+1,-1,-1,-1,+1,-1) \nonumber \\
{\bf S}^3 &=&  (+1,+1,-1,-1,-1,+1,-1,-1,-1,+1,+1,-1,-1,-1,-1,-1,-1,-1,-1,-1) \nonumber \\
{\bf S}^4 &=&  (+1,+1,-1,-1,-1,+1,-1,+1,+1,+1,-1,-1,-1,-1,+1,+1,-1,-1,+1,-1) \nonumber \\
{\bf S}^5 &=&  (+1,+1,-1,-1,-1,+1,-1,+1,+1,+1,-1,-1,-1,-1,+1,+1,-1,-1,+1,-1) \nonumber \\
{\bf S}^6 &=&  (+1,+1,+1,-1,-1,+1,-1,-1,-1,-1,+1,-1,+1,+1,+1,-1,-1,-1,+1,-1) \nonumber \\
{\bf S}^7 &=&  (+1,+1,-1,-1,-1,+1,-1,+1,+1,+1,+1,-1,+1,-1,+1,-1,+1,-1,+1,+1) \nonumber \\
{\bf S}^8 &=&  (+1,+1,-1,-1,-1,+1,-1,+1,+1,+1,+1,-1,+1,-1,+1,-1,+1,-1,-1,+1) \nonumber \\
{\bf S}^9 &=&   (-1,+1,+1,-1,-1,+1,-1,+1,-1,-1,+1,-1,+1,-1,-1,-1,-1,+1,+1,-1) \nonumber \\
{\bf S}^{10}&=& (-1,-1,-1,-1,-1,+1,-1,+1,-1,-1,-1,+1,+1,-1,-1,-1,-1,+1,+1,-1) \nonumber \\
{\bf S}^{11}&=& (-1,+1,-1,-1,-1,+1,-1,+1,+1,-1,+1,-1,-1,-1,+1,-1,+1,-1,-1,+1) \nonumber \\
{\bf S}^{12}&=& (-1,+1,-1,+1,-1,+1,-1,+1,+1,-1,+1,-1,+1,-1,+1,-1,+1,-1,-1,+1) \nonumber 
\end{eqnarray}
During  a transition  between one  q-ESS and  the next,  the
systems behaviour becomes  very hectic. Starting at (a)  we see that a
new mutant has invaded the previous coherent configuration (originally
similar to (b)) with a negative interaction with most of the existing
flowers but a strong enough  positive interaction with one of them for
it to  survive. This causes the  coherent state to  be destabilised. In
(b) we see that although the  new mutant does not survive for long, it
has  drastically reduced the  population at  ${\bf S}^1$, which  in
 turn  has a harmful effect  on ${\bf S}^2$ and ${\bf S}^3$.
 In  (c) we observe that  two further new mutants have been able to
 invade, this is due to the reduced fitness of
the  original sites from  the effect  of the  first invader.   The new
mutants are transient,  they represent steps on an  adaptive walk. The
system is now in a situation  where it is partly executing such a walk,
and partly still in the  previous coherent state.  This continues into
(d), where we can  see that ${\bf S}^1$ and ${\bf S}^3$ are still
  holding on, and their complete  first circles evince  they are
  still reproducing.   By (e),
however,  things have  changed  again. The  adaptive  walkers are  now
out-competing the originals,  ${\bf S}^1$ has become extinct and  
${\bf S}^3$ has a very
low population. We  also observe the formation of  a double flower 
(${\bf S}^7$ and ${\bf S}^8$)  which consists  of two fit  centres in adjacent  locations in
genome space. The adaptive walk continues  for some time until a
new q-ESS is found at (f).

We have studied the distribution of non-zero couplings 
${\bf J}({\bf S}^a,{\bf S}^b)$
between a given occupied position ${\bf S}^a$ and another occupied position
${\bf S}^b$. During the q-ESS this distribution is
narrow and its average is smaller than during the transition periods,
where the distribution broadens. 

\noindent{\bf Epoch Distributions}\\
It is interesting to take a further look at Fig. 5. One
notices that the distribution of lifetimes of occupied positions 
reaches as far out as the distribution of q-ESS durations.
In fact we observe in the simulations that positions sometimes
 are able to remain occupied
across the transition from one q-ESS to the following,
corresponding to  a species that survives a mass extinction.
Fig. 5 also shows clearly that the periods of hectic 
reconfiguration typically last for a significantly smaller
number\linebreak

\vspace{6.5cm}
\noindent of generations than do the q-ESS periods.
Finally, it is very interesting to note that both the
lifetimes of individual positions and the distribution of q-ESS epoch
lengths are power-law-like with exponents around -2.3 and
-1.8 respectively. We mention that the distribution of
q-ESS durations can be compared to the distribution of lifetimes of
genera obtained from the fossil record. The latter has a 
shape  similar to the distribution of q-ESS durations shown in Fig. 5.
Power-law fits to the fossil record data leads to an exponent 
around 2. For a recent analysis of data from the
fossil record  see Newman \& Sibani (1999).

\noindent{\bf Adaptation Level Increases}\\
We now turn to a discussion of the overall long time
effect of the dynamics of the TaNa model. How does
the genomic composition of early configurations differ
from those generated after hundreds of thousands of generations?

\begin{figure}[ht]
\vspace{0cm}
\centerline{\hspace{-0cm}
\psfig{figure=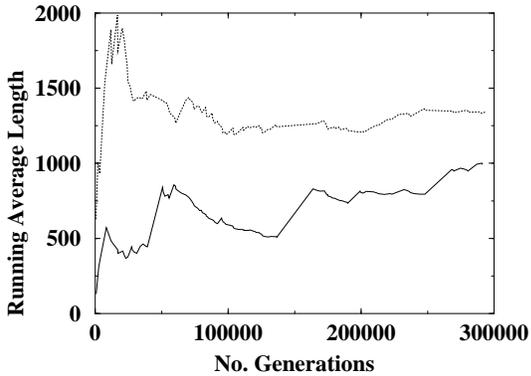,width=7cm,angle=-90}}
\vspace{0cm}
\caption{ 
Running Averages of the duration of q-ESS and transition periods. 
At the end of a particular state, we evaluate  the average
length of states up until that point. We see that the average
length of the  transition periods settles down and fluctuates slightly 
around constant, whereas the average q-ESS length continues to increase.
} 
\label{8}
\end{figure}

In Fig. 8 we show the running average of the durations of the q-ESS
as well as the transition epochs. One notices that there is no significant
trend in the duration of the hectic periods of rearrangement separating
the consecutive q-ESS. The average duration of the q-ESS periods, however, slowly increases with time. This means that the entire ecology gradually
becomes more stable. Or we may say that the ecology (represented by the
distribution of the population through genome space)  becomes increasingly better adapted; not adapted to some fixed external environment, but
adapted in the sense that the ecology as a whole achieves
collectively increasingly stable configurations among the total
set of all possible ways of distributing a population through genome
space. Does this mean that eventually some maximally ``fit'' or adapted
configuration is reached? Our simulations indicate, as expected,
that the time to reach a stationary state increases exponentially with
increasing genome length $L$. We will accordingly expect that for 
biologically
relevant systems an ecology would never have the time to reach a final
stationary state. Moreover, even if the system becomes stationary in
the sense that the average duration of the q-ESS becomes time independent,
switching between different equally well adapted configurations is likely
to continue forever. From the statistical mechanics of disordered systems
we do not expect the optimally adapted configuration to be unique.
Hence transitions between equally maximally adapted  configurations 
may continue even in the mathematical limit of infinitely long time.

The increase of the average duration of the q-ESS can be viewed as an
optimising process. This is in accordance with the suggestion (Mayr, 1988)
that the effect of biological evolution is to optimise some quantity.
The identification of the quantity being optimised is still debated
(Fogel \& Beyer, 2000). Unfortunately we cannot identify a specific 
mathematical function of the distribution $n({\bf S},t)$ in genome space
which is optimised as an effect of the dynamics. However,     
it is very interesting to relate the average duration of the q-ESS
to the extinction rate. Due to insufficient statistics we cannot,
unfortunately, make a quantitative comparison. We note, 
qualitatively, that an increasing average duration of the q-ESS corresponds
to a decreasing extinction rate. 
This is consistent with Raup \& Sepkoski's (1982)analysis of the fossil record,
which suggests
that the extinction rate might have declined through the Phanerozoic.

\subsection{Sexual Reproduction}
We now briefly discuss a model in which all individuals are assumed
to reproduce sexually. More detail will be the presented in a
future communication.

\noindent{\bf Long Time Behaviour}\\
In Fig. 9a and Fig. 9b, we show the temporal behaviour of the total number
of individuals together with the occupancy in genome space.

\begin{figure}[ht]
\vspace{-2cm}
\centerline{\hspace{-0cm}
\psfig{figure=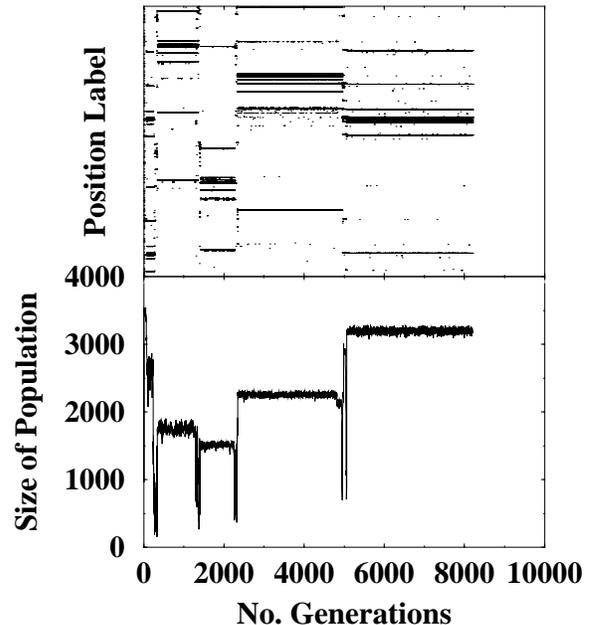,width=10cm,angle=0}}
\vspace{-2cm}
\caption{ 
The total number of individuals (below) and the occupation plot of
the species (above). The horizontal axis is the generational
time. The different plateaus in the population size correspond to 
rearrangements of the population in genome space. Parameters are
$L=20$, $\mu=0.07$, $p_{kill}=0.2$, $p_{mutate}=0.01$, $c=100$ and
$\theta=0.25.$ 
} 
\label{9}
\end{figure}
We have assumed that the maximum number of genes which
parents can differ, $d_{max}=2$ and in Fig. 9b, we plot only the
species occupancy, that is we have coarse grained genome space
with a resolution of $d_{max}$. This is done in the following way:
In each time step we identify the position with the largest population,
we lump this position together with all positions within a distance 
$d_{max }$. 

Next we find the position with the second largest population, and lump this location together with all positions within distance $d_{max}$. We continue
this until all occupied positions have been considered. 
The locations in genome space are labelled in a convenient, but
arbitrary, way. For each time step we place a dot along the y-axis
for each occupied (coarse grained) positions in genome space.  
Finally, along the x-axis, we convert time steps into time measured in
generations. We observe that, similar to the asexual case, the model
evolves through a set of q-ESS phases separated by short transition
periods. We also emphasise that well established species can
be identified as the well separated locations in genome space
where the population is concentrated.

\noindent{\bf Lifetime Statistics}\\
In Fig. 10, we show the distribution of lifetimes of
occupations of individual multiple occupied positions 
in genome space. A slow power-law-like
decay is observed. Note the similarity with the
distribution found in the asexual case and
with the distributions reported from the fossil record , see e.g.
Newman \& Sibani (1999).

\begin{figure}[ht]
\vspace{0cm}
\centerline{\hspace{0cm}
\psfig{figure=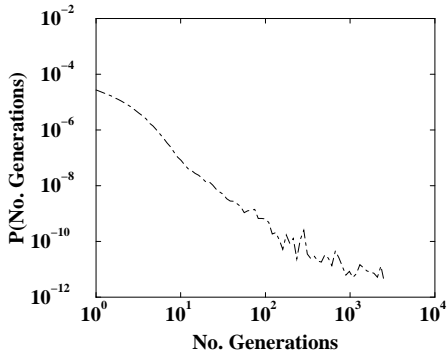,width=8.5cm,angle=-90}}
\vspace{0cm}
\caption{ 
Distribution of positions's lifetime (log-log plot) for the sexual
case. Same simulation as in Fig. 9.
} 
\label{10}
\end{figure}

\subsection{Competition Between Sexual and Asexual Populations}
To observe more directly the differences in behaviour between sexual and
asexual populations we have constructed a mixed reproductive mode
model. Here, individuals are given an extra gene which does not
explicitly enter the Hamiltonian, but instead dictates an
individual's reproductive mode. Mutations to this gene occur during
reproduction in the normal way. An asexual parent may potentially
 produce two sexual offspring 
whereas a sexual parent may produce at most one asexual offspring. 
This is compensated by assuming the
asexual mutation rate for the reproductive mode gene to be half that
for the sexual. In this way we eliminate any net drift
induced by the rates of mutation from one reproductive mode to the other.

In Fig. 11, we plot the total population of the system, along with
those of the two subpopulations (i.e. the numbers of sexual and
asexual individuals present at a given time) we see that in the coherent
phases the system is predominantly asexual, despite the large
fluctuations in population, whereas during coherent phases we see the
opposite: the population becomes predominantly sexual. The reason for
this effect is not clear, but we believe that the tight structures
(clusters of neighbour or nearest neighbour positions all of significant 
occupancy)  evolved in genome space by a sexual population may be
more suited to a constant environment than the more scattered ones
observed for asexual populations. 

This hypothesis is supported by the observation that a
sexual population with $d_{max}=L$ shows q-ESS  with asexual
dominance (see Fig. 11b). A sexual population of this sort
would not necessarily form such tight stable structures since
the pressure to have nearby mate-genomes of high fitness would
not be present.

The observed asymmetry between the asexual and the sexually
reproducing subpopulations for small values of $d_{max}$ may also 
be related time scales. The sexually reproducing individuals
need other individuals in the immediate vicinity of genome space
which takes time to establish. However, the mixing of ``genes''
involved in the recombination during sexual evolution allows 
the sexual population to scan through larger portions of genome
space, and therefore, if given time, this population is likely to
find the better adapted configurations.  

\begin{figure}[ht]
\vspace{0cm}
\centerline{\hspace{0cm}
\psfig{figure=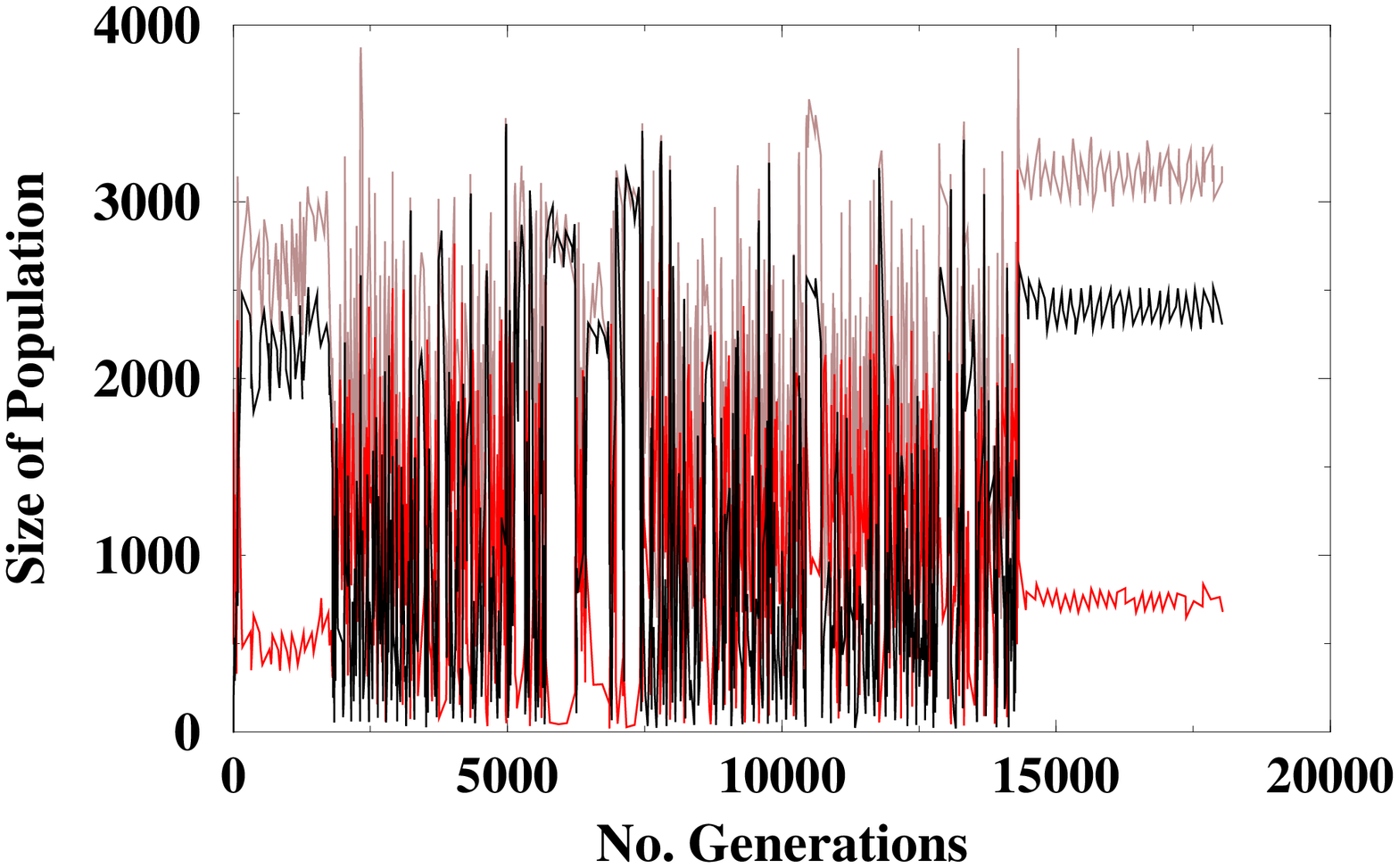,width=6cm,angle=-90}}
\vspace{0cm}
\label{11a}
\end{figure}

\begin{figure}[ht]
\vspace{0cm}
\centerline{\hspace{0cm}
\psfig{figure=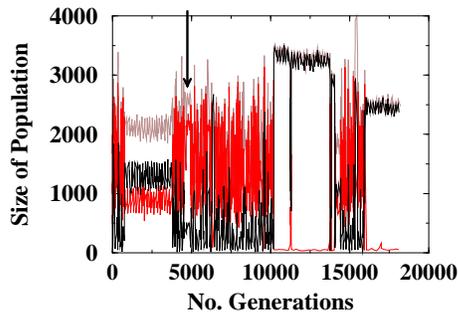,width=6cm,angle=0}}
\vspace{0cm}
\caption{ 
 Population Sizes versus Time in Mixed-Reproductive-Mode
Systems. Again we see a division in behaviours between q-ESS and
transition periods. (a) During q-ESS the sexual population
dominates, whereas the reverse is observed during transitions.
Evidence to support the hypothesis that this is due to tightly
packed flower structures evolved by a sexual population is supported by
(b) in which the selection pressure to form these structures is removed
by allowing all sexual individuals to reproduce. The arrow marks a
q-ESS in which the asexual population dominate. Others q-ESS
where the asexual population dominates are 
observed on shorter timescales than are visible on this plot.
} 
\label{111b}
\end{figure}

\section{Discussion and Conclusion}
We have discussed a simple, very general mathematical metaphor
(Gravrilets 1999) by which we can study the long time
(of order $10^5$ or $10^6$ generations) behaviour of an ecology.

Both asexual and sexually reproducing populations 
evolve through a set of relatively stable configurations,
the q-ESS, separated by short transition periods of hectic
reorganisation of the genomic composition of the ecology. The population
segregates in genome space into well separated clusters of highly occupied 
positions. Speciation events occur when a position or a tight cluster of
positions undergoes successive bifurcations in genome space. This type of behaviour is observed for a broad
range of control parameters.

The co-evolutionary dynamics produce a highly tangled interdependent
population of species. The evolution gradually increases the robustness
of the entire ecology against fluctuations in the genomic and physical environment. In agreement with analysis of the fossil record we
find that the average duration of the q-ESS increases slowly
with time.

\section{Acknowledgement}
We have benefited greatly from discussions with A. Burt, 
A.M. Lerio, R.R.B. Azevedo and, at a very early stage,
from conversations with C. Godfray.

\section{References}

\noindent Bak P. \& Sneppen K. (1993) 
{\it Punctuated Equilibrium and Criticality in a Simple Model
of Evolution}. Phys. Rev. Let. {\bf 71}, 4083-4086.

\noindent Bj\o rnstad, O.N., Sait, S.M., Stenseth, N.C., Thompson, D.J., and Begon, M. (2001). {\it The impact of specialized enemies on the 
dimensionality of host dynamics. }
Nature, {\bf 409}, 1001-1006.  

\noindent Darwin, C. (1859).
{\it On the Origin of Species by Means of Natural Selection}.
John Murray (this ed. 1951, Oxford University Press)

\noindent Doncaster, C.P.,  Pound, G.E., and Cox, S.J., (2000).
{\it The ecological cost of sex}, Nature {\bf 404}, 281-285.

\noindent Eigen, M., McCaskill, J. \& Schuster, P. (1988).
{\it Molecular Quasi-Species}. J. Phys. Chem. 92, 6881-6891.

\noindent Fogel, D.B. and Beyer, H.-G. (2000). {\it Do Evolutionary
Processes Minimize Expected Losses?} J. theor. Biol. {\bf 207},
117-123.

\noindent Gavrilets, S. (1999). 
{\it A dynamical theory of speciation on holey adaptive
 landscapes.} Am. Nat. 154, No.1, 1-22.

\noindent Higgs, P.G. \& Derrida, B. (1992).
{\it  Genetic distance and species 
formation in evolving populations.} 
 J. Mol. Evolution 35, No. 5, 454-465.

\noindent Kauffman, S. (1995). {\it At Home in the Universe}, Chap. 10 {\it
Coevolution}.
  Viking, London.

\noindent Maynard Smith, J., (1982). {\it Evolution and the theory of games}. Cambridge, Cambridge University Press.

\noindent Maynard Smith, J., (1995). {\it The Theory of
Evolution}, Cambridge University Press.

\noindent Mayr E., (1988). {\it Toward a New Philosophy of
Biology: Observations of an Evolutionist.} Harvard, MA: Belknap

\noindent Newman, M.E.J. \& Sibani P., (1999). 
{\it Extinction, diversity and survivorship of 
taxa in the fossil record.} Proc. R. Soc. Lond. B {\bf 266}, 
1593-1599. 

\noindent Pimm, S.L., (1991).
 {\it The Balance of Nature}, University of Chicago Press.

\noindent Raup, D.M. \& Sepkoski Jr., J.J. (1982) {\it Mass
extinction in marine fossil record.} Science {\bf 215}, 1501-1503.

\noindent Wagner, H., Baake, E. \& Gerische, T. (1998).
{\it Ising Quantum Chain and Sequence Evolution}. J. Stat. Phys. 92, 1017-1052.

\end{document}